\documentclass[10pt, emptycopyrightspace]{ewsn-proc}
\usepackage{pifont}
\usepackage{paralist}
\usepackage[usernames,dvipsnames,svgnames,table]{xcolor}

\usepackage{color}
\usepackage{colortbl}
\newcommand{\cmark}{\color{ForestGreen} {\ding{51}}}%
\newcommand{\xmark}{\color{BrickRed}{\ding{55}}}%

\usepackage{xspace}
\usepackage[hidelinks]{hyperref}
\usepackage{graphicx}
\usepackage{balance}
\usepackage{comment}
\usepackage{tabularx}
\usepackage{adjustbox}
\newcommand{\numdevices}{11\xspace}
\newcommand{\numexp}{462\xspace}

\numberofauthors{3}

\author{
\alignauthor Yang Li \\
    \affaddr{University College London}\\
    \email{yang-li.22@ucl.ac.uk}
\alignauthor Anna Maria Mandalari \\
    \affaddr{University College London}\\
    \email{a.mandalari@ucl.ac.uk} \\
\alignauthor Isabel Straw \\
    \affaddr{University College London}\\
    \email{isabel.straw.20@ucl.ac.uk}
}

\title{Who Let the Smart Toaster Hack the House? An Investigation into the Security Vulnerabilities of Consumer IoT Devices}

\begin{document}

\maketitle

\begin{abstract}
For smart homes to be safe homes, they must be designed with security in mind. Yet, despite the widespread proliferation of connected digital technologies in the home environment, there is a lack of research evaluating the security vulnerabilities and potential risks present within these systems. Our research presents a comprehensive methodology for conducting systematic IoT security attacks, intercepting network traffic and evaluating the security risks of smart home devices. We perform hundreds of automated experiments using \numdevices popular commercial IoT devices when deployed in a testbed, exposed to a series of real deployed attacks (flooding, port scanning and OS scanning). Our findings indicate that these devices are vulnerable to security attacks and our results are relevant to the security research community, device engineers and the users who rely on these technologies in their daily lives. 
\end{abstract}
\section{Introduction}

  \label{sec:intro}
In the age of technology, our homes are changing. The rapid development of smart devices and the emergence of the Internet of Things (IoT) is reshaping the environment in which we live and the means by which we carry out our daily lives \cite{IoT_review}\cite{number_of_IoT_devices}. The IoT has been described as the ubiquitous network of devices which communicate with one another, without human interaction, permeating the infrastructure of our experience \cite{IoT_challenges}. The smart home is an application of an IoT environment, which comprises the physical entities and connectivity present in domestic settings \cite{IoT_challenges}. In the home, the first primitive IoT device was a remotely controllable toaster, introduced in 1990 as a proof-of-concept \cite{glossary_iot}. Since this time consumer IoT devices for the home have flooded the market, ranging from smart TVs, to connected light bulbs, thermostats and door locks \cite{glossary_iot}. According to an extrapolation from 2021, the number of IoT devices will rise to 51 billion in 2023 and continue to increase in the foreseeable future, estimated to reach 75 billion by 2025 \cite{glossary_iot}\cite{number_of_IoT_devices}. 


Security is of paramount concern to smart home users, evidenced by research from Aldossari and Sidorova, who highlighted the relationship between consumer device acceptance, trust and notions of security and privacy \cite{number_of_IoT_devices}. Traditionally in IT security, domain goals have consisted of ensuring confidentiality, integrity and accountability of systems and messages \cite{schiller2022landscape}. Yet researchers have illustrated that such frameworks are outdated and fail to account for the evolving risks of IoT systems \cite{mandalari2021blocking}. IoT systems operate 24/7 and therefore are always available for attacks (e.g. botnet attacks)~\cite{kolias2017ddos} and the heterogeneous plethora of possible devices present within the system novel security issues \cite{IoT_challenges, mandalari2021blocking}.

 The harm that can result from an attack is dependent on the end-point functionality of the device, which in the smart home encompasses a spectrum of harms ranging from a faulty smart fridge to an unresponsive smoke detector \cite{IoT_challenges}. Previous research has described the means by which smart home design can open up the door to risks that range from exposing the privacy of householders, to facilitating crimes such as burglary using video feeds, to tampering with healthcare appliances to enact physical harm \cite{hammi2022survey}.


The rising prevalence of IoT devices results in a growing range of security and privacy risks. Many IoT devices can involuntarily become part of a botnet \cite{ddos_iot} and may be vulnerable to Denial of Service (DoS) attacks \cite{IoT_dos}. Other risks include leakage of Personally Identifiable Information (PII) because of lack of encryption or authorisation, misactivation \cite{ren2019information}, or malware attacks \cite{IoT_malware}. In this paper, we aim to evaluate the security and privacy risks of consumer IoT devices by developing a methodology for conducting systematic IoT attacks and intercepting the devices' network traffic.  We use our large-scale  IoT testbed, along with several Raspberry Pi 4s (RPi 4), to launch over \numexp automated experiments against \numdevices IoT devices. 
The security attacks, privacy threats, and vulnerabilities that we evaluate include network attacks (e.g., port scanning, flooding, \emph{etc}.).

Surprisingly, we find that IoT devices are indeed vulnerable to well-known and documented IoT security attacks. 

Our key research contributions include the following:
\begin{compactitem}
\item We develop an automated methodology for evaluating security vulnerabilities in common consumer IoT devices using large-scale, diverse experiments and sets of attacks;
\item We assess the security vulnerabilities of popular IoT devices against existing network and device attacks, and identify privacy risks.
\end{compactitem}

In summary, we find that consumer IoT devices are highly vulnerable to common IoT security attacks. We argue for increased security and privacy in this space, given the risks for the users when IoT devices are compromised.  

We make our experiment software and datasets available at \url{https://github.com/SafeNetIoT/spices}.

\section{Assumptions}

In this section, we summarize the threat model and goals of this work.

\subsection{Threat Model}
We consider the following threat model. 

\noindent \emph{Adversary.} The adversary is any party that can access the internal IoT device network, such as malicious IoT devices. 

\noindent \emph{Victim.} The victim is any person in a smart home that owns an IoT device in a smart home.

\noindent \emph{Threat.} We assume the presence of malicious or compromised IoT devices in a smart home. 
The malicious device has access to the home router.
Adversaries may be incentivized to compromise other devices in the network for inferring user activities or denying the usage of them.
We consider security threats (Mirai, Scan, \emph{etc}.~\cite{doshi2018machine,antonakakis2017understanding}).

\subsection{Goals and Non-Goals}

The main goal of this work is to analyze the reaction of consumer IoT devices to common security threats. In particular, this work answers the following research questions (RQ):

\noindent \emph{RQ1. Are consumer IoT devices vulnerable to common security attacks?} 
Our goal is to characterize how IoT devices react to security attacks. To address this, we propose a testbed for systematically studying their reaction and capturing their network traffic.

\noindent \emph{RQ2. Do the IoT devices detect threats?}  
IoT devices may have security protection techniques in place and notify the user or manufacturer when detecting security threats. We check their capability to do so.  

\noindent \textbf{Non-Goals.}
In this initial study, we do not consider the following as goals, and leave them for future work.

\noindent \emph{No control over how an IoT device works internally.}
We consider the IoT device as a blackbox, we do not have control over how an IoT device works internally. However, we have the capability to interact with them using their companion app, and we can track their network activity.

\noindent \emph{We do not test all threats.}
Our methodology only focuses on a subset of security threats for
every IoT device, so that we can cover the same threats for different devices. 

\noindent \emph{Consumer IoT devices.}
We focus on IoT devices that target consumers; we do not consider medical or industrial IoT devices.

\section{Testbed}
\label{sec:testbed}

 \begin{figure}[t!]
  \centering
  \includegraphics[width=0.9\columnwidth]{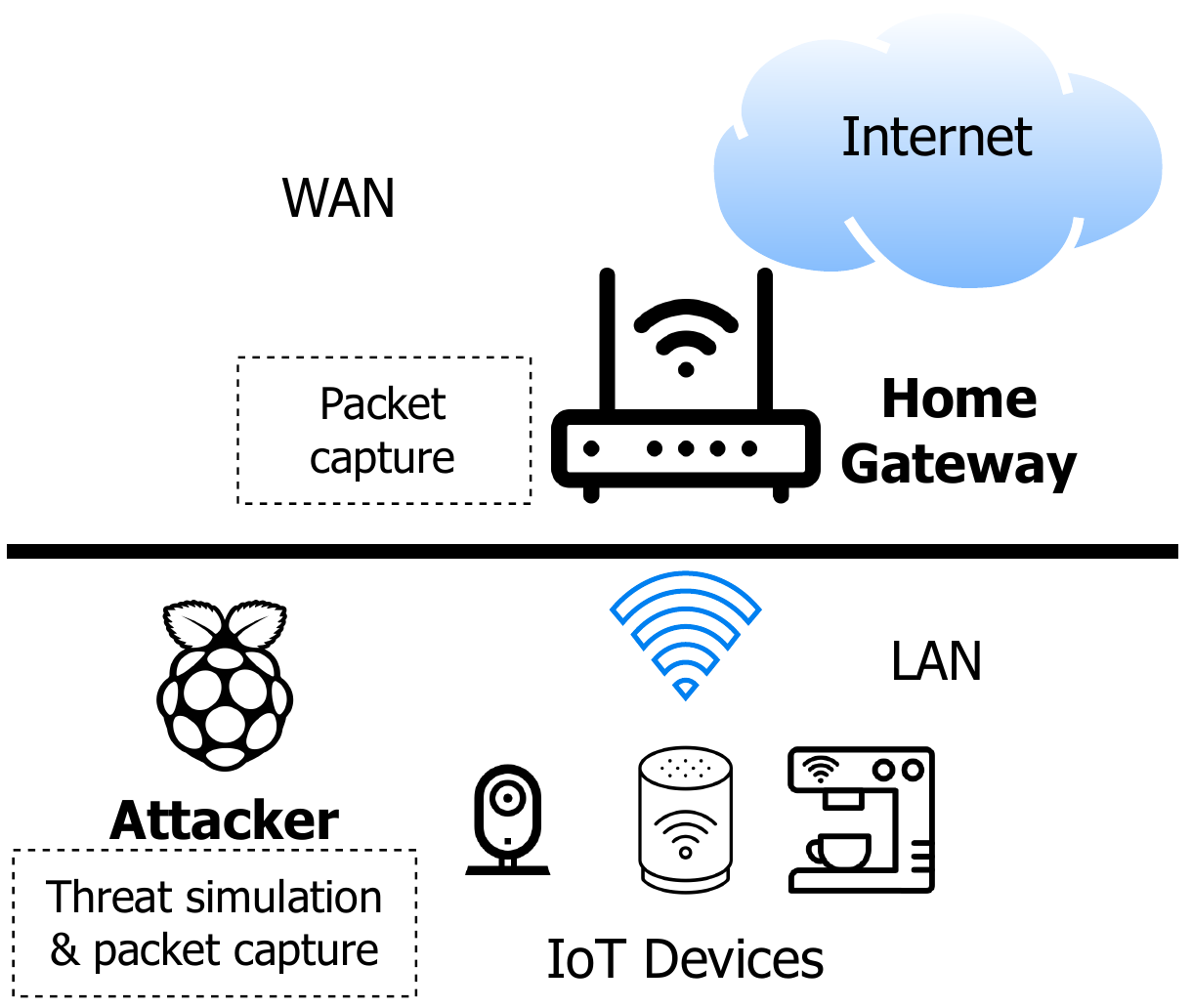}
  \vskip 0.1in
  \caption{Overview of the testbed.}
  \label{fig:setup}
\end{figure}

In order to have a controlled environment for threat emulation, we build the testbed shown in Figure~\ref{fig:setup}. The testbed  consists of: (\emph{i}) a \emph{gateway} that provides IP connectivity to the IoT devices from the Internet and has the capability of capturing all the network traffic of the IoT devices; (\emph{ii}) the \emph{Attacker}, an RPi which acts as an IoT device in the network, (\emph{iii}) the \emph{IoT devices under test}, a group of popular IoT devices all connected to the gateway; (\emph{iv}) \emph{threat scripts}, run at the attacker to execute IoT security threats experiments.
More details on each component are presented below.

\subsection{Gateway}
The gateway is configured using a NAT setup. It has two network interfaces, a WAN interface with a public IPv4 address, and a LAN interface with a private IP address, used to give NAT Internet connectivity to the IoT devices and the attacker. The gateway has DHCP capabilities, effectively trying to mimic the typical configuration of a smart home network.
The gateway is also capable of intercepting all the network traffic from the IoT devices and the attacker using tcpdump.
The gateway is physically connected to an Android phone via the Android Debug Bridge, and has the ability to control the IoT devices through their Android companion app. Each device's traffic is filtered by MAC address into separate files.

\subsection{Attacker}
This component is an RPi, with one network interface (Wi-Fi) connected to the gateway, where all the IoT devices under test are connected.
The attacker is responsible for running the \emph{threat scripts} to simulate threats originating from an IoT device in the LAN network.

\subsection{IoT Devices}
\label{sec:iot}
Table~\ref{tbl:tested_devices} shows the IoT devices we consider. We examine \numdevices consumer IoT devices typically deployed in a smart home. We select these devices to provide diversity within different categories and among the most popular ones we could find on the market. We choose devices in 4 categories: smart speakers (3), smart doorbells (2), smart cameras (3), and appliances (3). 
To better represent how IoT devices behave, we try to keep their default configuration and privacy settings unaltered, and we do not perform user-initiated firmware upgrades. 
All the IoT devices are connected to the Wi-Fi interface of the gateway, making them part of the LAN where the attacker is located (i.e., their private IP addresses and DNS servers are assigned by the DHCP server of the gateway).
Their network traffic is captured by the gateway.

\begin{table}[t!]
\centering
\caption{IoT devices.}
\vskip 0.1in
\begin{tabular}{|l|l|}
\hline
Category       & Device                            \\ \hline
Smart speaker  & Bose Smart Speaker 500            \\ \cline{2-2} 
               & Sonos One (Gen2)                  \\ \cline{2-2} 
               & Echo Dot 5                        \\ \hline
Smart doorbell & Ring Chime Pro                    \\ \cline{2-2} 
               & Ring Video Doorbell (2nd Gen)     \\ \hline
Smart camera   & Google Nest Cam                   \\ \cline{2-2} 
               & SimpliSafe Security Camera Indoor \\ \cline{2-2} 
               & Furbo 360° Dog Camera             \\ \hline
Appliances     & WeeKett Smart Wi-Fi Kettle        \\ \cline{2-2} 
               & Govee Alexa LED Strip Lights      \\ \cline{2-2} 
               & Sensibo Sky Smart AC              \\ \hline
\end{tabular}
\vskip -0.1in
\label{tbl:tested_devices}
\end{table}

\subsection{Threat Scripts}
We simulate security threats programmatically using \emph{threat simulation scripts}, which, depending on the type of threat, are run on the attacker (threats originating targeting the IoT devices under test).
We generate three threats involving Denial of Service (DoS), port scanning, and OS scanning. 
We make our threat scripts available at \url{https://github.com/SafeNetIoT/spices}.

\section{Methodology}

In this section, we report the methodology we use for answering our research questions.
We propose an experimental setup that detects the IoT devices' traffic and semi-automatically classifies our attacks as successful or not. 

We evaluate devices' defensive measures against various attacks and compare within and across categories.
\subsection{Assessing Device Vulnerability}

\subsubsection{Attacks Definition}
We define a list of attacks that can be simulated in a testing environment to assess devices' vulnerabilities. The attacks include SYN (port 80), UDP, DNS, and fragmented IP flooding, as well as port scanning and OS scanning~\cite{mandalari2023protected}. The flooding attacks and scanning attacks are implemented in separate and configurable scripts. We run the scripts on an RPi 4, connected to the same LAN as the IoT devices under attack.
Our threat script uses Nmap~\cite{lyon2008nmap} to launch port and OS scan attacks and tcpreplay~\cite{tcpreplay} for flooding attacks. Each type of flooding attack is repeated continuously ten times to allow sufficient time for the device to detect and mitigate such an attack. We also check for inconsistency in device behaviours across repeated attacks.

\subsubsection{Attacks Validation}
In order to assess whether the attacks are able to reach their targets (IoT devices) as expected, we set up a second RPi 4 connected to the same LAN as the first one.  We conduct attacks targeting the second RPi and perform packet capture on it. We then verify that simulated attacks reach the second RPi thus validating our experiment.

\subsection{Assessing Device Reaction}
We run tcpdump continuously on the gateway to capture network packets for all devices connected. All flooding and scanning network traffic is also captured and separated into different folders per device.

We then use Tshark~\cite{tsoukalos2015using} to analyse the packet captures. We implement detection scripts for filtering all the packets coming from the attacker. By applying the filter, we are able to intercept the corresponding replies to the simulated attacks.

We determine whether an attack is successful by analysing the target device's reaction. If the device implements countermeasures that detect ongoing attacks and mitigate the consequences of the attacks, the attack is considered to be unsuccessful, even if the device's normal functionality is interrupted (e.g. interrupted video streaming to its companion application). However, the attack is considered successful if no defensive measures can be observed on the target device's captured traffic and the device's normal activity is halted. 
\section{Evaluation}
We now answer our research questions by identifying and characterizing the reaction of IoT devices to security threats. 

\subsection{Flooding}

\begin{table}[]
\caption{Devices and flooding attacks (Successful attack: {\cmark}, Unsuccessful attack: {\xmark}).}
\vskip 0.1in
\centering
\begin{tabular}{|l|c|c|c|c|}
\hline
Devices                           & SYN  & UDP  & DNS  & Frag. IP \\ \hline
Bose Speaker            & {\cmark}    & {\xmark}         &{\xmark}    & {\cmark}                   \\ \hline
Sonos One (Gen2)                  & {\xmark}         & {\xmark}         & {\xmark}         & {\cmark}                   \\ \hline
Echo Dot 5                        & {\xmark}         & {\xmark}         & {\xmark}         & {\cmark}                   \\ \hline
Ring Chime Pro                    & {\xmark}         & {\xmark}         & {\xmark}         & {\cmark}                   \\ \hline
Ring Doorbell                     & {\xmark}         & {\xmark}         & {\cmark}         & {\cmark}                   \\ \hline
Google Nest Cam                   & {\xmark}         & {\xmark}         &{\xmark}         & {\cmark}                   \\ \hline
SimpliSafe Cam            & {\xmark}         & {\xmark}         & {\xmark}         & {\cmark}                   \\ \hline
Furbo Camera             & {\xmark}         & {\xmark}         & {\xmark}         & {\cmark}                   \\ \hline
WeeKett Kettle        & {\xmark}         & {\cmark}         & {\cmark}         & {\cmark}                   \\ \hline
Govee Lights      & {\xmark}         & {\xmark}         & {\cmark}         & {\xmark}                   \\ \hline
Sensibo Sky              & {\xmark}         & {\cmark}         & {\cmark}         & {\cmark}                   \\ \hline
\end{tabular}
\vskip -0.1in
\label{tbl:flood}
\end{table}

\begin{table}[t!]
\centering
\caption{Devices and identified open ports (filtered ports are reported in \textcolor{ForestGreen}{green}).}
\vskip 0.1in
\begin{tabular}{|p{2.5cm}|p{5cm}|} \hline
Device&Identified Open Ports\\ \hline
Bose Speaker& 80/7000/8082/8083/8085 /8091/8200/30030/40002 /40031/40035 \\ \hline
Sonos One & 1400/1410/1443/1843/7000 \\ \hline
Echo Dot 5 & 1080/4070/8888/55442/55443 \\ \hline
Ring Chime Pro & \textcolor{ForestGreen}{847/1003/1020/1393/3736/7240 /8173/12302/15986/16891 /17704/17944/17993/18682/20307 /21257/23825/24669/25781/25958 /25997/26757/27234/28363/29161 /32466/33377/33544/33616/33862 /35470/38657/44100/46108/46194 /47199/50852/51212/52663/54739 /55524/55530/56621/65488}\\ \hline
Ring Doorbell & Blocking ping probes \& none found \\ \hline
Google Nest Cam & 8012/10101/11095 \\ \hline
SimpliSafe Cam & 19531\\ \hline
Furbo Camera & None found \\ \hline
WeeKett Kettle& 6668 \\ \hline
Govee Lights & None found \\ \hline
Sensibo Sky & None found \\ 
\hline\end{tabular}
\vskip -0.1in
\label{tbl:device_and_ports}
\end{table}

Table~\ref{tbl:flood} shows the (un)successful rate of attacks for each device.  
During SYN flooding, Bose Smart Speaker 500 replies to the SYN packets with SYN/ACK packets. Other devices, except Echo Dot 5, reply to every SYN packet with RST/ACK. Among all devices, Bose Smart Speaker 500 performs the worst in SYN DoS attacks as it replies to inbound SYN with SYN/ACK packets, which would hold the corresponding communication ports half-open, consuming the most resources and making the device stop working. On the contrary, other devices, excluding Echo Dot 5, defend themselves against SYN flooding by resetting those half-open connections, reducing unnecessary resource consumption caused by the attack. 

In UDP flooding, Sensibo Sky Smart AC and the Weekett kettle reply with ICMP port unreachable packets with significant delay. Other devices, excluding Echo Dot 5, only reply to a fraction of messages with significant delay. The ICMP port unreachable messages are error messages indicating that the requested UDP port is unavailable or closed~\cite{postel1981internet}. Due to the connectionless nature of the UDP protocol, UDP flooding can successfully render a device unusable without establishing two-way conversations. Hence, all devices perform more or less the same against UDP flooding attacks, as their normal communications are reduced or halted when attacks are in progress.

During DNS flooding, Ring Video Doorbell (2nd Gen), the Weekett kettle, Govee Alexa LED Strip Lights, and Sensibo Sky Smart AC also reply with ICMP port unreachable messages. The rest of the devices reply sparsely, not including the Echo Dot 5. This can be due to their limited resources or designed defensive mechanisms to mitigate the effects of DNS flooding attacks. We conclude that it is challenging to assess devices' performance under DNS DoS attacks without having access to the devices' source code.

Under the IP fragmentation attack, none of the tested devices responds, except the Govee Alexa LED Strip Lights, which replies with an ICMP message stating Time-to-live exceeded (Fragment assembly time exceeded). This indicates that the Strip Light is designed to discard or drop the fragmented packets when it takes too long to assemble them into a complete IP packet. Other devices might have different defensive designs that do not involve sending such ICMP packets.

The Bose speaker still sends application data during SYN, UDP, and DNS flooding but stops working during the IP fragmentation attack. The Ring Chime Pro pings its server during those flooding but also stops working during the IP fragmentation attack. The devices could be sending those messages to potentially report the ongoing attacks or seek assistance during flooding events. Other devices' normal communications with their server are interrupted during the flooding. After the attack, they resume communicating with their servers.

Echo Dot 5 does not respond to any of the attacking packets, which may indicate better security practices. No inconsistency can be identified between repeated attacks.

\subsection{Port Scanning}
The port scanning results identify no open ports on Furbo Camera, Govee Lights, and Sensibo Sky. The Ring Doorbell blocks the ping probes, so in this case, the attack is not successful. Various numbers of open ports are identified on the rest of the devices, as shown in Table \ref{tbl:device_and_ports}. Ports 80 to 8200 on the Bose speaker are associated with known services, contrarily to ports 30030-40035. Although the Ring Chime Pro has the largest number of open ports, they are all shown in \textit{filtered} state, meaning Nmap cannot determine whether they are open. The identified open ports on other devices are in \textit{open} state. An open port is actively listening for incoming connections and suggests that a service or application is running on that port. A filtered port indicates that some form of filtering or blocking mechanism is in place, which could indicate the presence of a firewall. It is worth noting that a port that is closed during scanning could open up if an application uses it.



\subsection{OS Scanning}
\label{subsec:os_scanning}
All tested smart speaker devices have Linux as OS, as shown in Table \ref{tbl:os_scanning_results}.  Interestingly, the scanning results show that the Ring Video Doorbell is likely to have a similar OS to the 2N Helios IP VoIP doorbell. There are devices whose OSes cannot be identified. Those devices could have defensive mechanisms like network filtering (like the Ring Chime Pro) or obfuscation. If not, it could be due to the limitations of the scanning tool or too many similarities between OSes. The rest two appliances all utilise lightweight IP stacks as they are open-source and resource-efficient.

\begin{table}[t!]
\centering
\caption{Devices and identified Operating Systems (OS).}
\vskip 0.1in
\begin{tabular}{|l|p{5cm}|} \hline
\label{tbl:os_scanning_results}
Device&Operating System\\ \hline
Bose Speaker & Linux 3.2 - 4.9\\ \hline
Sonos One (Gen2) & Linux 3.2 - 4.9 \\ \hline
Echo Dot 5 & No exact match, can be Linux \\ \hline
Ring Chime Pro & Too many fingerprints match\\ \hline
Ring Doorbell & 2N Helios IP VoIP doorbell (95\%) \\ \hline
Google Nest Cam & Too many fingerprints match \\ \hline
SimpliSafe Cam & Too many fingerprints match\\ \hline
Furbo Camera & Too many fingerprints match \\ \hline
WeeKett Kettle & No exact OS matches\\ \hline
Govee Lights & Espressif esp8266 firmware (lwIP stack), NodeMCU firmware (lwIP stack) \\ \hline
Sensibo Sky & Philips Hue Bridge (lwIP 1.4.1), Philips Hue Bridge (lwIP stack) \\
\hline\end{tabular}
\vskip -0.1in
\end{table}
\section{Discussion}

Our findings demonstrate vulnerabilities across a range of consumer technologies. We now turn to consider the impact this may have on the user in their lived environment, the limitations of our methodology and ethical considerations. 

\noindent \textbf{User Implications.}
The harm posed by a security threat is contextual to the role of the device in the environment. For example, malfunctioning smart heating systems may be more consequential than a compromised kettle. The security flaws we demonstrate in lighting systems (LED Strip Lights) and sound systems (e.g. smart speakers) illustrate that adversarial attacks may significantly impact the sensory experience of an occupant in the home. Flooding attacks that result in DoS may render a system unresponsive, for example preventing an occupant from activating their lighting system (an outcome that could be particularly distressing at night and if imposed for criminal intent, e.g. burglary). Further research is needed to explore whether these attacks pose a risk to smart lock systems, which, if successful, could prevent an individual from entering/exiting their property. The manipulation of lighting systems is a heightened concern for photosensitive individuals, such as epilepsy sufferers, who have been identified as at risk of seizures from security attacks on smart home lighting systems \cite{Epilepsy}. Furthermore, domestic violence researchers have exposed concerning trends in interpersonal abuse, reporting that perpetrators have exploited smart light and sound systems to inflict physical and psychological harm on victims \cite{Tech_Abuse1, Tech_Abuse2}. The additional success of flooding attacks on appliances, such as kettles, illustrated that these simple methods could disrupt the ability of an occupant to use the equipment within their home.  

Beyond flooding attacks, our work exposes open ports present within connected systems, raising the question of possible exploits that may be enacted through attacks on these pathways. It is possible that with these ports being open, they may be accessed remotely, allowing an adversary to take control of a device. Unfortunately, we are unable to determine the current use of these ports and the means by which they may be manipulated. We leave this as future work. However, their open state allows us to question the harms that could result from attacks aimed at these targets. In particular, the open ports present in Ring doorbells and smart speakers raise the question of whether adversaries could transmit audio into a living environment and impose incessant sound signals. While we found no open ports in the Smart Air Conditioning (AC) machinery, further research is required to explore the risks of exploitation in the range of these devices. 

\noindent \textbf{Privacy and Security Implications.}
Attacks on sensory systems are likely to be immediately apparent to the occupant who is disturbed by these manipulations of the environment. Other exploits, such as privacy attacks, maybe more surreptitious. For example, we have demonstrated vulnerabilities in smart security cameras. If the video footage from these devices is inconspicuously obtained, the data may be shared elsewhere, resulting in a significant breach of occupant privacy and regulation. The implications of our research should therefore be considered through the General Data Protection Regulation (GDPR) framework, which sets the standard for data protection and privacy in the EU and the European Economic Area. For developers, there is an extent of literature that explores the application of GDPR’s governing principles and provisions to IoT infrastructure ~\cite{GDPR}. 

The European Guidelines developed for IoT security are relevant to our findings. The framework states that if a port is not being used, that port should be closed, yet our findings demonstrate a plethora of open ports in consumer home technologies for unclear reasons. While ENISA and NIST guidelines~\cite{ENISA-IoT, nist} have been developed to improve design practices and secure the supply chain of IoT, they are currently not mandatory, and we need a methodology for understanding their compliance. Additional research has proposed solutions at the edge for protecting the user from IoT attacks \cite{mandalari2021blocking}.

\noindent \textbf{Limitations.}
Our exploration of security vulnerabilities in the smart home infrastructure is constrained by a number of limitations. Firstly, these devices have been examined as black-boxes, in which no attempt has been made to reverse engineer their code or response strategies (as these resources are often unavailable or undocumented). Furthermore, our experiments are limited to \numdevices devices which form only a small proportion of the vast and ever-growing consumer IoT market. In the evolving IoT space, new devices (with potentially new vulnerabilities) are constantly appearing on the market. 

\noindent \textbf{Ethical Considerations.}
In our experiments, we do not cause any real threat on the Internet. All experiments are contained within our own testbed. When conducting the experiments, we fully respected the ethical guidelines defined by our affiliated organization.
\section {Related Work}

Many works have assessed the security and privacy risks of consumer IoT devices. Approaches used during assessments include running simulated attacks~\cite{simulated_attacks, sivanathan2017experimental, mandalari2023protected}, static source code analysis \cite{source_analysis}, network traffic interception \cite{ren2019information, 7815045}, and binary code reverse engineering \cite{binary_RE}. However, their methodology does not allow them to run experiments that assess security threat reactions automatically. Running simulated attacks and network traffic interception were chosen for their scalability regarding the number of devices that can be tested simultaneously and the black-box nature of many devices. 

Babun et al.~\cite{babun2021survey} perform an analysis of popular smart home platforms. The authors focus on commercial and open-source platforms, pointing out their limitations when dealing with IoT data and apps.
In contrast, our study is about IoT security threats, offering a comprehensive tool for assessing their reactions to common attacks.
\section{Conclusion}
\label{sec:conclusion}

Detecting security threats on smart home IoT devices is an important ongoing challenge.
Commercial IoT devices are appearing in the market and being offered by different vendors, but there has been no insight into how they react to security threats.

In this paper, we took a quantitative approach in auditing some of the IoT devices available in the market, as well as analyzed their reaction to common security threats.
We developed a scalable and automated methodology for evaluating the effectiveness of these attacks against known IoT devices.
Our evaluations using 3 security threats on 4 device categories on an advanced IoT testbed indicate underwhelming protection for commercially available IoT devices. 
They often are vulnerable to common security attacks; further, they do not include any security protection or user alerting system. 

Based on our findings, we argue there is a need for IoT security and privacy systems deployed specifically for IoT devices and developed at the edge. To assist with such efforts, we make our datasets (IoT devices packet captures) and software public to encourage the creation of such systems and better security compliance from IoT vendors at \url{https://github.com/SafeNetIoT/spices}. We will maintain the codebase regularly to keep it up-to-date.



\balance
\bibliographystyle{unsrt}
\bibliography{ewsn-workshops}  
\end{document}